\documentclass{article}
\usepackage[utf8]{inputenc}
\usepackage{spconf,amsmath,graphicx,amsfonts}
\usepackage{tikz,booktabs,float,color,adjustbox,multirow,tabularx}
\usepackage{hyperref}
\usepackage[noadjust]{cite}
\usepackage{tikz}
\usetikzlibrary{arrows}
\title{Tie Your Embeddings Down: Cross-Modal Latent Spaces for End-to-end Spoken Language Understanding}
\name{\begin{tabular}{c}Bhuvan Agrawal\thanks{$^\dagger$ Work done during author's internship at Amazon Alexa}$^\dagger$, Markus M{\"u}ller, Martin Radfar, Samridhi Choudhary\\ Athanasios Mouchtaris, Siegfried Kunzmann\end{tabular}}
\address{Alexa Machine Learning, Amazon}
\begin{document}
\ninept
\maketitle
\begin{abstract}
End-to-end (E2E) spoken language understanding (SLU) systems can infer the semantics of a spoken utterance directly from an audio signal. However, training an E2E system remains a challenge, largely due to the scarcity of paired audio-semantics data. In this paper, we treat an E2E system as a multi-modal model, with audio and text functioning as its two modalities, and use a cross-modal latent space (CMLS) architecture, where a shared latent space is learned between the `acoustic' and `text' embeddings. We propose using different multi-modal losses to explicitly guide the acoustic embeddings to be closer to the text embeddings, obtained from a semantically powerful pre-trained BERT model. We train the CMLS model on two publicly available E2E datasets, across different cross-modal losses and show that our proposed \textit{triplet loss} function achieves the best performance. It achieves a relative improvement of $1.4\%$ and $4\%$ respectively over an E2E model without a cross-modal space and a relative improvement of $0.7\%$ and $1\%$ over a previously published CMLS model using $L_2$ loss. The gains are higher for a smaller, more complicated E2E dataset, demonstrating the efficacy of using an efficient cross-modal loss function, especially when there is limited E2E training data available.
\end{abstract}

\section{Introduction}
Spoken language understanding (SLU) is the task of inferring the semantics of user-spoken utterances and is the core technology behind voice assistant systems like Alexa, Google Home, Siri and more.
The traditional approach to SLU uses two distinct components to sequentially process a spoken utterance: an automatic speech recognition (ASR) model that transcribes the speech to a text transcript, followed by a natural language understanding (NLU) model that predicts the domain, intent and entities given the transcript \cite{lee2010recent}.
However, this modular design admits a major drawback.
The two component models (ASR and NLU) are trained independently with separate objectives.
Errors encountered in the training of either model do not inform the other.
For example, for an input utterance `turn on the light', if the ASR hypothesis is missing a single word, the word error rate (WER) is $25\%$.
However, if the missing word is `on' or `light', the NLU model will produce an incorrect prediction with a high probability.
On the other hand, the information needed by the NLU model to make a correct prediction will still be present if the missing word is `the'.
Yet, the WER (the error metric the ASR is optimized for) does not differentiate this effect when looking at the final model performance.
 error metric the ASR is optimized for) does not reflect this effect on the NLU downstream component.

An increasingly popular approach to address this problem is to employ models that predict SLU output directly from a speech signal input \cite{martinInterspeech2020,haghani2018audio,serdyuk2018towards,lugosch2019speech,chen2018spoken,qian2017exploring,ghannay2018end}.
This class of models, also referred to as speech-to-interpretation (S2I) models, are trained in an end-to-end (E2E) fashion to maximize the SLU prediction accuracy.
Such models typically have smaller footprints than their modular counterparts, making them attractive candidates for performing SLU in resource constrained environments like devices on the edge.
However, availability of sufficient quantity of good quality speech data with associated semantic labels is key to achieving comparable performance to the traditional, cascaded counterparts.
A paucity of such datasets becomes a major bottleneck for these SLU systems.

A variety of methods, like knowledge transfer, data augmentation and component-wise pre-training have been explored to address this issue.
For example, curriculum and transfer learning strategies are used in \cite{caubriere2019curriculum, tomashenko2019investigating} to gradually fine-tune the SLU model on increasingly relevant datasets, followed by finally training it on low-resource in-domain data.
Authors in \cite{lugosch2019speech, bhosale2019end} leverage the large amount of transcribed speech data to pre-train an ASR model on phoneme and word-level targets.
This is then used to initialize the first few layers of the SLU model, that is fine-tuned on a smaller training set containing paired transcripts and SLU labels.
This method leads to a near perfect accuracy on their datasets.
However, a principled study performed by the authors in \cite{mckenna2020semantic}, revealed that most of these methods report high performance on SLU datasets of relatively low semantic complexity, often representing targeted SLU use-cases.
As the complexity of the dataset increases and the SLU task becomes more sophisticated, the performance of these models starts degrading.

Authors in \cite{huang2020leveraging} attempt to address this problem by combining component pre-training, knowledge transfer and data augmentation approaches to create a robust SLU model.
They employ the successful multi-modal learning approach of unifying the latent representation space across different modalities \cite{ngiam2011multimodal}, and utilize the fact that the SLU model is essentially a multi-modal model, with speech and text functioning as its two modalities.
One of the main advantages of this architecture is the common latent space that allows leveraging both acoustic and text-only data for model training.
This cross-modal feature space leads to a better representation and gives the model a robust inductive bias for a semantically complex SLU task.
BERT-based text embeddings \cite{devlin2018bert} are used to `guide' acoustic embeddings.
An $L_{2}$ loss between the text and  acoustic embeddings is used to explicitly tie the cross-modal latent representation space, leading to better intent classification accuracy.

In this work, we further explore the multi-modal view of SLU models by experimenting with different approaches to learn a robust cross-modal latent space.
Specifically, we experiment with (i) different loss functions to tie the acoustic and text embeddings; and (ii) using adversarial training methods on the E2E SLU model.
We try two other loss functions apart from the $L_{2}$ loss -- pairwise ranking loss and triplet loss.
These loss functions have shown to be more effective, than the $L_{2}$ loss, in learning joint feature representations in the regimen of image processing for learning tied image-sentence representations \cite{kiros2014unifying,frome2013devise,faghri2017vse,wang2017adversarial}.
Our goal is to have the encoders generate embeddings in the same latent space, so that the origin of the embeddings becomes indistinguishable.
A method alternative to the previous losses is the use of adversarial training, where a classifier is trained to detect the modality of a given embedding and forces the (audio) encoder to output acoustic embeddings that can fool this classifier, pushing them to be closer to the structure of textual embeddings.

We train the so-called cross-modal latent space (CMLS) E2E SLU model architecture defined in \cite{huang2020leveraging} on two publicly available SLU datasets \cite{lugosch2019speech, saade2018spoken} and study the effect of these cross-modal embedding losses and of the adversarial training on the intent accuracy of the model.
We show that the triplet loss has the best performance across both these datasets.
It achieves a relative improvement of $1.4\%$ and $4\%$ respectively\footnote{For the dataset in \cite{lugosch2019speech} and \cite{saade2018spoken} in order.} over an E2E model without a cross-modal space and a relative improvement of $0.7\%$ and $1\%$ over the model from \cite{huang2020leveraging} that uses the $L_2$ loss. A visual inspection of the acoustic embeddings via a t-SNE plot, reveals that the embeddings learned using the best performing CMLS loss (\textit{triplet loss}) are able to better separate the utterances belonging to different intents, even if they are acoustically similar, when compared to the embeddings from a model that does not have a CMLS setup.

To the best of our knowledge, this is the first attempt to systematically apply losses that have shown success in learning cross-modal representations in relatively mature textual-visual or speech-visual multi-modal domains, to SLU models for improving the SLU tasks.
We hypothesize that using an appropriate loss function and cross-modal training methodology is key to achieving tighter coupling and hence better performance, as the SLU tasks get more complex.

In the Section \ref{sec:method} we detail our methodology where we explain the CMLS model architecture that we implemented and the different embedding losses used .
This is followed by our experimental setup and model configurations used for training in Section \ref{sec:experiments}.
We also discuss the results in Section \ref{sec:experiments} across the two datasets and perform some analyses on the differences in the cross-modal representations learned across the different loss functions. We conclude with key findings and future directions of this work in Section \ref{sec:conclusion}.

\begin{figure*}[!ht]
	\centering
	\includegraphics[trim=0cm 1.5cm 0cm 0cm,width=0.75\textwidth]{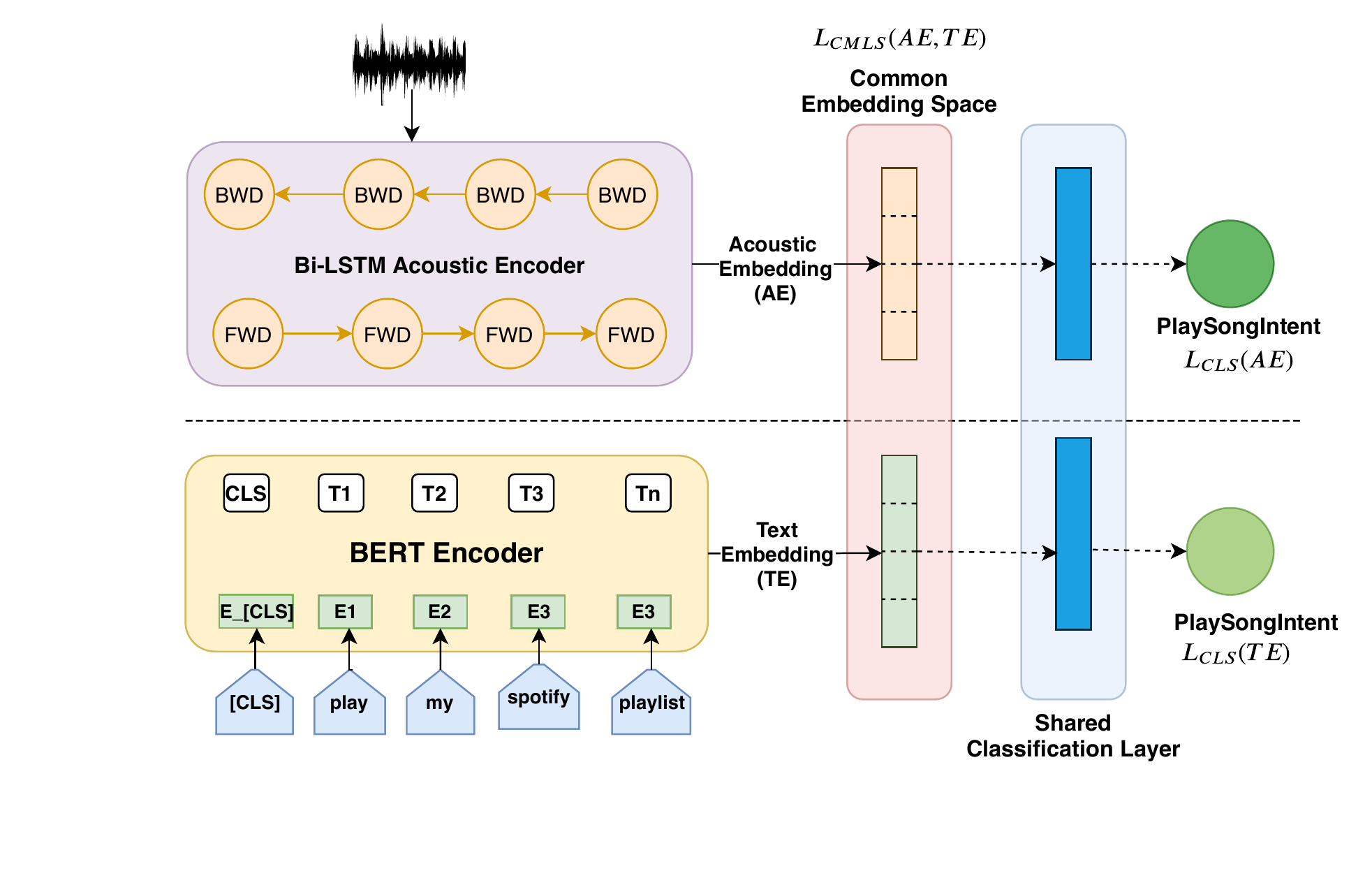}
	\caption{The cross-modal latent spaces (CMLS) SLU model architecture. $L_{CE}(AE/TE) $ is the cross-entropy classification loss for the acoustic/text embedding, whereas, $L_{CMLS}(AE,TE)$ is one of our cross-modal embedding loss to bring the acoustic and text embeddings together. The text-encoder component below the dotted line is removed for inference}
	\label{fig:systemarch}
\end{figure*}
\section{Cross-Modal Latent Spaces}
\label{sec:method}

A major challenge in building an efficient E2E SLU system is the availability of appropriate datasets.
While there is a plethora of ASR and NLU specific datasets available, the is a scarcity of good quality E2E SLU datasets that have paired utterance audio and semantic labels, to allow for an end-to-end training of the model.
An approach to overcome this, is building a system that has a tied space for multiple modalities. This presents the advantage of leveraging different, modality specific datasets that can help achieve better generalization for the final SLU task.
The tied cross-modal space can be achieved by unifying representations across modalities. An example would be, projecting the acoustic and text-only data into the same embedding space, and learning the parameters of this joint space to optimize the SLU task-loss.
This enables us to train an SLU model in an end-to-end fashion on the smaller E2E dataset, but also leverage the large amount of ASR- or text-only datasets to learn a robust latent space for the final task.

A first solution to learning such a cross-modal latent space \textit{(CMLS)} between the acoustic and text modalities in an E2E SLU model was proposed in \cite{huang2020leveraging}.
The `text' modality latent space is represented by using the encoded representation of an utterance text from a pre-trained BERT model. Whereas, the `acoustic' modality is represented by using a multi-layer Bi-LSTM model to create an acoustic embedding of the utterance audio.
Along with the task-specific classification loss\footnote{A cross-entropy loss on intent classification is used.}, they use an $L_{2}$ loss between the text and  acoustic embeddings, to explicitly tie the cross-modal space.
A shared classification layer is jointly trained on both the acoustic embedding (AE) and text embedding (TE).

While this method of an explicit embedding loss, added to the task-specific loss, led to an improved performance over the baseline model, there are more effective ways of mapping representations from different modalities into a common space.
In the regimen of computer vision, an established approach is to map embeddings of different modalities (e.g. images and captions) into the same embedding space using the pair wise ranking loss or the triplet loss.
While the $L_2$ loss forces the embeddings to approximate each other in a canonical way using the distance, we rely on alternative multi-modal losses to build a stronger CMLS model in this work. We use the same E2E SLU model setup as in \cite{huang2020leveraging}, but use three different losses to learn a robust CMLS model.
In the next subsection \ref{sec:method:subsec:arch}, we outline our model architecture and describe each loss in detail in subsection \ref{sec:method:subsec:losses}, followed by subsection \ref{sec:method:subsec:adversarial} which addresses our approach on adversarial training.

\subsection{Model Architecture}
\label{sec:method:subsec:arch}
Our proposed CMLS model consists of three sub-modules as shown in Figure \ref{fig:systemarch}: an acoustic encoder, a BERT encoder, and a shared classification layer.
The acoustic encoder is a multi-layer Bi-LSTM network that computes the acoustic embedding (AE) from the acoustic features of the utterance audio. This is done by max-pooling the last layer Bi-LSTM states across the time dimension to obtain a fixed-dimensional vector that summarizes the input audio, independent of the length of the input signal.
In order to obtain the text embedding (TE) of the input utterance, we use a pre-trained BERT model that takes the  utterance text as an input. As is common in BERT-based encoders, the last layer transformer-encoder representation of the [CLS] token is used as the text-embedding of the utterance.
The shared classification layer is a fully-connected network, followed by a softmax to predict the semantic label (intent in our case). It produces an intent prediction using both the AE and TE separately, resulting in computing an embedding specific classification loss $L_{CLS}(AE/TE)$  as shown in Figure \ref{fig:systemarch}.

Explicit loss to tie the AE and TE is specified by $L_{CMLS}$, that can be one of the three losses that we define in the following section.
The discriminator network for the adversarial training (not shown in the Figure) is composed of linear layers.
Depending on the dataset, we optimized the number of layers, units per layer and the weight of the adversarial loss.

We back-propagate the embedding and SLU task losses only to the acoustic branch. This is due to the limited amount of training data available, where the BERT model would overfit because of its size. As mentioned in more detail later in section \ref{sub:sec:combined_loss}, the joint loss seen by the acoustic branch is constructed by combining AE specific classification loss and one of the aforementioned embedding loss ($L_{CMLS}$).
The BERT-based text pipeline is only used during training to guide the AEs and is discarded for inference. This architecture has the advantage of being easily extensible to support more modalities during training while at the same time keeping the resource footprint constrained during inference.
The CMLS model is not limited to using a BERT model, the method is agnostic to the type of embeddings used.
\subsection{Embedding losses}
\label{sec:method:subsec:losses}
We evaluate three different loss functions ($L_2$ loss, pairwise ranking loss and triplet loss) to tie the embeddings originating from two modalities together into a common latent space.
\subsubsection{$L_2$ loss}
Introduced in \cite{huang2020leveraging} to tie the latent space, the $L_2$ loss is computed as follows:
\begin{equation}
\label{eq:l2loss}
\mathcal{L}_{E}(\mathbf{x}_1, \mathbf{x}_2) = d(\mathbf{x}_1, \mathbf{x}_2) = || \mathbf{x}_1 - \mathbf{x}_2 ||_2^2
\end{equation}
where $\mathbf{x}_1$ denotes the acoustic embedding and $\mathbf{x}_2$ denotes the text embedding of the same utterance. $d(\mathbf{x}_1, \mathbf{x}_2)$ expresses the distance between $\mathbf{x}_1$ and $\mathbf{x}_2$.

\subsubsection{Pairwise Ranking Loss}
Moving beyond the $L_2$ loss, the pairwise ranking loss \cite{li2017improving} allows us to train the network using the following loss formulation:
\begin{equation}
\label{eq:pair}
\mathcal{L}_{E}(\mathbf{x}_1, \mathbf{x}_2, t) = td(\mathbf{x}_1, \mathbf{x}_2) + (1-t)\max\{0, m - d(\mathbf{x}_1, \mathbf{x}_2)\}
\end{equation}
where, $\mathbf{x}_1$ denotes the acoustic embedding and $\mathbf{x}_2$ denotes the text embedding.
$\mathbf{x}_1$ and $\mathbf{x}_2$ need not necessarily represent the same utterance. $t$ is a binary variable indicating if $\mathbf{x}_1$ and $\mathbf{x}_2$ have the same intent ($t=1$) or a different one ($t=0$) and $m$ is the margin which controls the minimum distance between the negative pairs.
This is a tunable hyperparameter.

If both samples belong to the same intent, the loss forces them to be closer together (similar to the $L_2$ loss), but if they originate from different intents, then the loss is $\max\{0, m - d(\mathbf{x}_1, \mathbf{x}_2)\}$ and this pushes the embeddings further apart.
\subsubsection{Triplet Loss}
In contrast to other losses, the triplet loss \cite{chechik2009large,schroff2015facenet} uses three examples: the current training example, called anchor $\mathbf{x}_a$ and  $\mathbf{x}_+$, $\mathbf{x}_-$ are positive and negative examples respectively. Formula \ref{eq:triplet} shows how the loss is computed.
\begin{equation}
\label{eq:triplet}
\mathcal{L}_{E}(\mathbf{x}_a, \mathbf{x}_+, \mathbf{x}_-) = \max\{0, m + d(\mathbf{x}_a, \mathbf{x}_+) - d(\mathbf{x}_a, \mathbf{x}_-)\}
\end{equation}
In our case, the anchor is the acoustic embedding and the positive example is the text embedding of an utterance with the same intent.
Similarly, the negative example is the text embedding of an utterance with a different intent.
For example, the anchor could be the acoustic embedding of `could you please increase the brightness' and the positive example could be the text embedding of `it's too dark in here' which both have the \texttt{IncreaseBrightness} intent.
The negative example, on the other hand, could be the text embedding of `change the lights to green' which has the \texttt{ChangeColor} intent.

The triplet loss aims at minimizing the intra-class variance while maximizing the inter class variance: samples within the same class are forced to be closer together in the feature space, while samples from different classes are pushed further apart.
Figure \ref{fig:triplet} shows embeddings in the latent space.
Using the triplet loss, the acoustic embeddings are shifted to minimize the distance between semantically similar words, while also maximizing the distance of semantically different words, even if they are acoustically close.
\tikzstyle{reddot}=[draw=red,fill=red]
\tikzstyle{bluedot}=[draw=blue,fill=blue]
\tikzstyle{singlearrows}=[black,line width=0.5mm,fill=white,preaction={-triangle 90,thin,draw,shorten >=-1mm}]
\tikzstyle{myarrows}=[black,line width=0.5mm,fill=white,preaction={-triangle 90,thin,draw,shorten >=-1mm}]
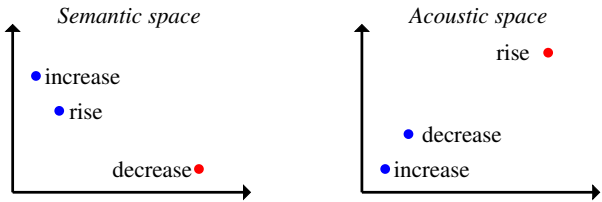
\begin{figure}[h]
\centering
\resizebox{0.44\textwidth}{!}{%
\begin{tikzpicture}[scale=0.425]
\draw[singlearrows] (0,0) -- (0,7);
\draw[singlearrows] (0,0) -- (10,0);
\draw[bluedot,line width=0.5mm] (1,5) circle (0.15);
\draw[bluedot,line width=0.5mm] (2,3.5) circle (0.15);
\draw[reddot,line width=0.5mm] (8,1) circle (0.15);
\node[align=center,font=\large,rotate=0] at (3.0,5) {increase};
\node[align=center,font=\large,rotate=0] at (3.15,3.5) {rise};
\node[align=center,font=\large,rotate=0] at (6.0,1) {decrease};
\node[align=center,font=\large,rotate=0] at (5,7.5) {\emph{Semantic space}};

\draw[singlearrows] (15,0) -- (15,7);
\draw[singlearrows] (15,0) -- (25,0);
\draw[bluedot,line width=0.5mm] (16,1) circle (0.15);
\draw[bluedot,line width=0.5mm] (17,2.5) circle (0.15);
\draw[reddot,line width=0.5mm] (23,6) circle (0.15);
\node[align=center,font=\large,rotate=0] at (18.0,1) {increase};
\node[align=center,font=\large,rotate=0] at (19.3,2.5) {decrease};
\node[align=center,font=\large,rotate=0] at (21.5,6) {rise};
\node[align=center,font=\large,rotate=0] at (20,7.5) {\emph{Acoustic space}};

\end{tikzpicture}
}
\caption{Leveraging the triplet loss in CMLS task: In contrast to previously used losses ($L_2$ loss), the triplet loss not only minimizes the distance between semantically related words but also maximizes the distance between those target words that are acoustically close but semantically opposite.}
\label{fig:triplet}
\end{figure}
\subsection{Combination of the Losses} \label{sub:sec:combined_loss}
The entire model is trained jointly and the total loss can be written as shown in Formula \ref{eq:combine}, where $\mathcal{L}_{cls}$ denotes the classification loss and $\mathcal{L}_{E}$ denotes the embedding loss.
$\lambda_1$ and $\lambda_2$ are hyperparameters that control the weights of the different losses in the sum.
\begin{equation}
\label{eq:combine}
\mathcal{L} = \mathcal{L}_{cls}^{acoustic} + \lambda_1\mathcal{L}_{cls}^{text} + \lambda_2\mathcal{L}_{E}
\end{equation}

\subsection{Adversarial Training}
\label{sec:method:subsec:adversarial}
A popular approach for adversarial training are GANs (generative adversarial networks) \cite{NIPS2014_5423}, where two networks are trained in tandem.
One (the generator network) is trained to generate/alter data samples and a second one (the discriminator network) is trained to classify whether the samples are generated by the generator or if they are `real' samples.
The generator is trained to fool the discriminator about the origin of the samples.
For CMLS, we face a similar scenario where we have multiple sources of samples while the objective is to generate embeddings which are invariant w.r.t. the source (here: acoustics or text).
We therefore use a discriminator network for adversarial training.
This network tries to predict if the corresponding embedding originates from the audio domain or the text domain.
Mathematically, we have a neural network $D$ that takes as input the embedding $\mathbf{x}$ from the common embedding space and outputs the probability that the embedding is from the text domain.
We assign a target of $1$ to the text modality and $0$ to the acoustic modality.
We denote the acoustic encoder by $AE$ and the BERT model by $TE$. The full adversarial objective can be written as in Formula \ref{eq:adversarial}.
$\theta$ denotes the parameters of the respective networks.
\begin{equation}
\label{eq:adversarial}
\min_{\theta_AE, \theta_TE} \max_{\theta_D} \mathbb{E}_{\mathbf{x}}[\log D(TE(\mathbf{x})) + \log (1 - D(AE(\mathbf{x})))]
\end{equation}

\section{Experiments and Results}
\label{sec:experiments}

In this section, we describe the datasets, model parameters, training configurations, and analyze our results\footnote{Source code is available at \url{https://github.com/alexa/alexa-end-to-end-slu}}.

\subsection{Datasets}
We use two publicly available SLU datasets to train and evaluate our model -- Fluent Speech Commands (FSC) \cite{lugosch2019speech} and Snips SLU \textit{SmartLights} \cite{saade2018spoken}. Both these datasets contain utterance text, corresponding audio and a semantic class label.

FSC is one of the largest public SLU datasets, containing $\approx 30,000$ utterances. Each utterance text is associated with a triplet - action, object and location. This triplet, functions as the `intent' label of the utterance, and becomes our target semantic class to be predicted. There are a total of 31 unique intent classes in the dataset.
Snips is a smaller SLU dataset, making the prediction task challenging. It contains $\approx 3,000$ utterances and 6 intent classes. Since this dataset does not have its own train/test/validation splits, we created a 80-10-10 split for train, validation and test respectively. 

We train two separate models corresponding to each dataset and study the effect of the losses and adversarial training for each of these datasets.

\subsection{Feature Extraction}
A Hamming window of the length 25 ms with frame rate of 10 ms is used to segment the audio files.
We used the \texttt{torchaudio} library to compute 40-dimensional MFCC features from the audio files directly.
We set the number of triangular Mel-frequency bins to 80.

\subsection{Model Training and Hyper-Parameters}
\label{sec:experiments:subsec:training}
The CMLS-SLU model described in section \ref{sec:method:subsec:arch}, consists of a 4-layer and 3-layer Bi-LSTM acoustic encoder for FSC and Snips respectively, with 512 hidden units per layer. The text-encoder comprises of a pre-trained BERT model\footnote{We use the BERT-base-cased models.} from \cite{wolf2019huggingface}, where the encoded representation of the [CLS] token from the last encoder layer is used as the text-embedding of the utterance.
The shared classification layer has an input size of 768 (the size of the embeddings) and the number of outputs depends on the dataset (31 for FSC and 6 for Snips SLU).

The cross-modal coupling between the acoustic and the text embeddings is achieved by using the three different losses defined in section \ref{sec:method:subsec:losses}. For each loss type, we train a new CMLS model and evaluate the results on the corresponding test split per dataset. A thing to note here is that the CMLS model trained on an $L_2$ embedding loss is the same as proposed in \cite{huang2020leveraging} and is therefore used as a basis of comparison for us to see the effectiveness of better cross-modal losses.

In order to show the efficacy of tying the text-acoustic cross-modal embedding spaces, we also train a \emph{baseline model} that just contains the multi-layer Bi-LSTM acoustic encoder, with a fully connected layer to perform intent classification, without any text encoder present. Furthermore, we also note the intent accuracy results noted by the authors of the FSC in their work \cite{lugosch2019speech} to compare the performance of the CMLS models on this dataset. Since we do not pre-train the acoustic layers of our model, we use their non-pretrained model results for comparison.

We perform a hyper-parameter search across different parameter combinations and use early stopping on the validation loss to select the best performing model for each embedding loss type.
We applied Adam optimizer \cite{kingma2014adam} with $\beta_1$ = 0.9, $\beta_2$= 0.98, and $\epsilon$ = 1e-9.  The learning rates for the acoustic encoder and BERT were set to 1e-3 and 2e-5, respectively for FSC which were decayed when the loss on the validation set plateaued.  For Snips we used the 1cycle learning rate policy with the max learning rate set to 6e-3.
For the adversarial generator network, we also performed a hyper-parameter search to tune the parameters for each dataset.
The parameters are listed in Table \ref{tab:hypparams} and differ between the two datasets.
\begin{table} [h!]
\centering
\begin{tabular}{lccc} \toprule
\textbf{Dataset} & \textbf{\# units} & \textbf{\# layers} & \textbf{loss weight}\\ \midrule
FSC & 256 & 2 & 0.1 \\
Snips SLU & 512 & 1 & 0.3 \\ \bottomrule
\end{tabular}
\caption{Discriminator network hyperparameters for adversarial training}
\vspace{-1em}
\label{tab:hypparams}
\end{table}

\subsection{Results}
We first present our results using different loss functions for each dataset and conclude with results of our adversarial experiments.
\vspace{-1em}
\subsubsection{Fluent Speech Commands}
The results on the FSC dataset are shown in Table \ref{tab:results-fsc}. We train three different CMLS SLU models on FSC, one per loss type. All the loss types are able to perform better (or equal to, in the case of ranking loss) than the baseline model. Triplet loss achieves the best performance with a relative improvement of $1.4\%$ over the baseline model and $0.4\%$ relative over the $L_2$ loss model from \cite{huang2020leveraging}.
When compared to the original accuracy reported on the FSC dataset by the authors in \cite{lugosch2019speech} for their non-pre-trained model ($96.6\%$), we see that the triplet loss achieves the best performance here too with a relative improvement of $1\%$. The $L_2$ loss is also able to beat the reported accuracy from \cite{lugosch2019speech} by $0.7\%$ relative. Both these results indicate the importance of having a cross-modal latent space, using a pre-trained text encoder in an E2E SLU model.

\vspace{-1em}

\begin{figure*}[ht!]
\centering
\begin{minipage}[b]{0.45\linewidth}
\includegraphics[width=0.95\textwidth]{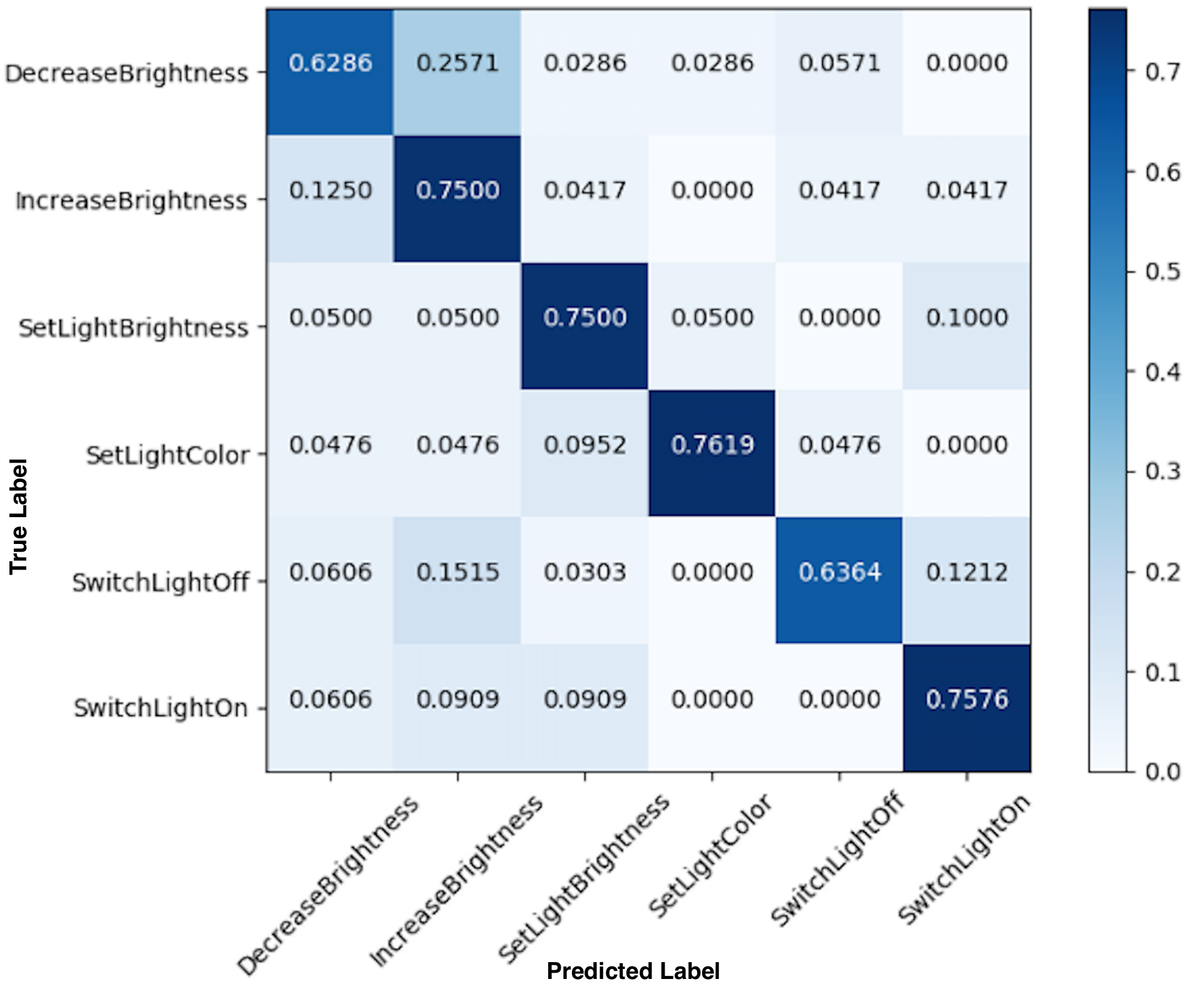}
\caption{\textbf{Baseline Model} - The confusion matrix of the predicted intents.}
\label{fig:conf1}
\hfill
\end{minipage}
\quad
\begin{minipage}[b]{0.45\linewidth}
\includegraphics[width=0.95\textwidth]{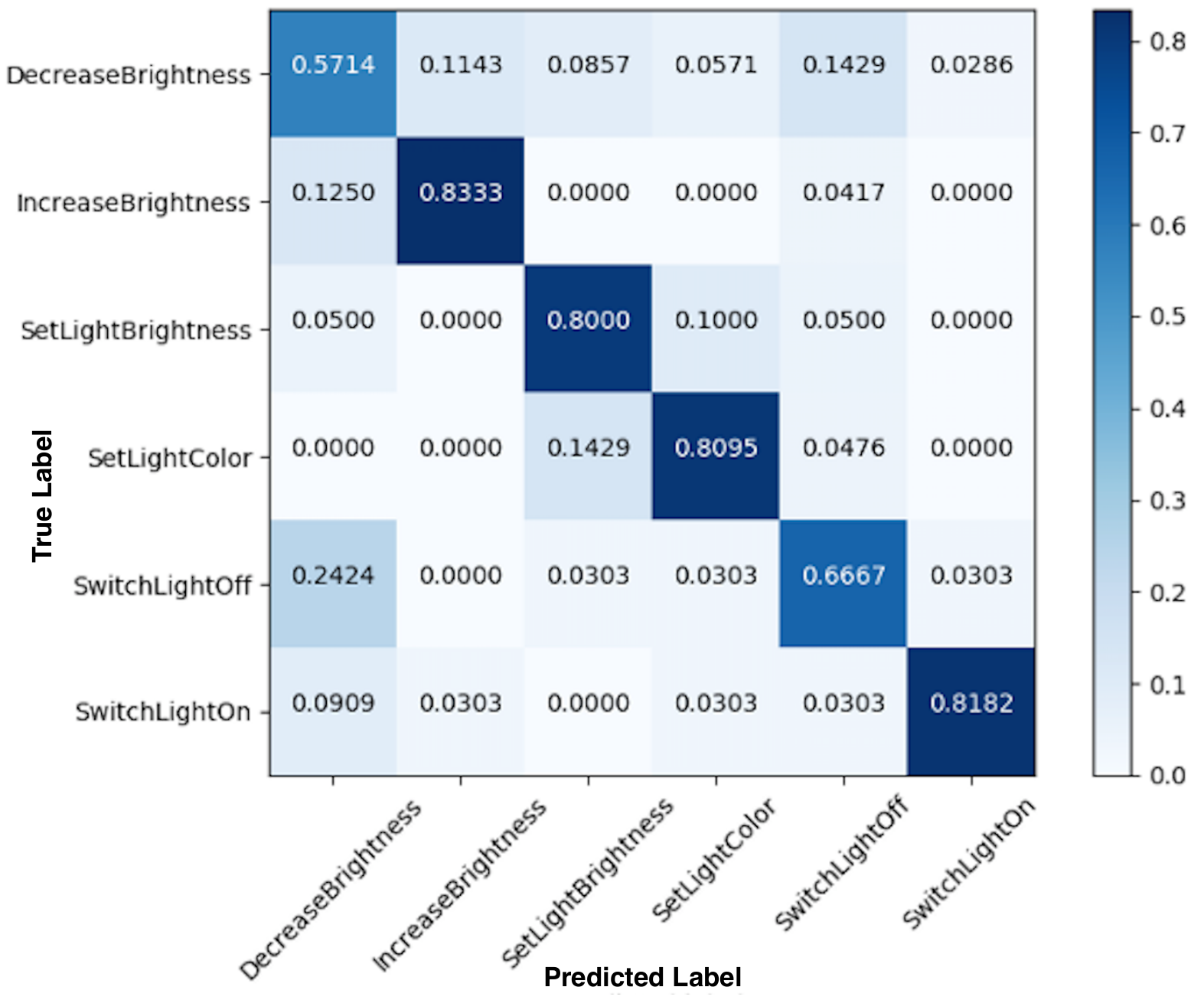}
\caption{\textbf{CMLS Model} - The confusion matrix of the  predicted intents when the triplet loss function is used.}
\label{fig:conf2}
\hfill
\end{minipage}
\end{figure*}

\begin{table}
\centering
\begin{tabular}{@{}lr@{}} \toprule
\textbf{Model} & \textbf{Accuracy (\%)}\\ \midrule
Baseline (Only acoustic, no CMLS) & 96.31\\
CMLS - $L_2$ loss (model in \cite{huang2020leveraging}) & 97.31 \\ \midrule
CMLS - Pairwise ranking loss & 96.31 \\
\textbf{CMLS - Triplet Loss }& \textbf{97.65} \\ \bottomrule
\end{tabular}
\caption{Intent accuracy on FSC. The baseline model is just the acoustic encoder with no Cross-Modal Latent Space (CMLS) while the CMLS model with $L_2$ loss is similar to the model proposed in \cite{huang2020leveraging}.}
\vspace{-1em}
\label{tab:results-fsc}
\end{table}
\subsubsection{Snips SLU}
We repeat the experiments on the Snips SLU dataset, the results of which are shown in Table \ref{tab:results-snips}. Similar to our previous observation, the CMLS SLU model is able to beat the performance of the baseline model, for all the embedding loss types, achieving a relative improvement of $3\%$ on average. The triplet loss has the best performance on this dataset too, with a relative improvement of $4\%$ over the baseline and $1\%$ over the $L_2$ loss model from \cite{huang2020leveraging}. Since this dataset is a more challenging dataset, owing to its small size, these performance improvements are notable. Moreover, these results also indicate that, agnostic of the SLU dataset used, having an explicit signal to learn a robust cross-modal embedding space, using a semantically powerful text-encoder as a guide, is important to achieve a robust SLU performance.

\begin{table}[ht!]
\centering
\begin{tabular}{@{}lr@{}} \toprule
\textbf{Model} & \textbf{Accuracy (\%)}\\ \midrule
Baseline (Only acoustic, no CMLS) & 70.84\\
CMLS - $L_2$ loss  (model in \cite{huang2020leveraging}) & 72.89 \\ \midrule
CMLS - Pairwise ranking loss & 72.29 \\
\textbf{CMLS - Triplet Loss} & \textbf{74.10}\\\bottomrule
\end{tabular}
\caption{Intent accuracy on the Snips SLU. The baseline model is just the acoustic encoder with no Cross-Modal Latent Space (CMLS) while the CMLS model with $L_2$ loss is similar to the model proposed in \cite{huang2020leveraging}.}
\label{tab:results-snips}
\end{table}
\begin{figure*}[!htbp]
\centering
\begin{minipage}[b]{0.45\linewidth}
\includegraphics[width=0.95\textwidth]{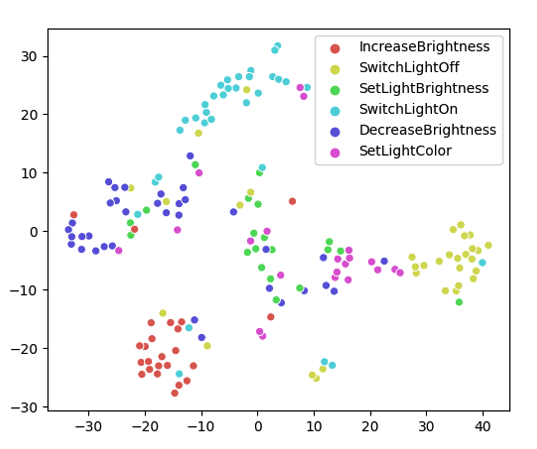}
\caption{\textbf{Baseline Model} - The t-SNE scatter plot of output intents. The intents clusters are not well-separated and have high-variance .}
\label{fig:tsnetriplet1}
\end{minipage}
\quad
\begin{minipage}[b]{0.45\linewidth}
\includegraphics[width=0.95\textwidth]{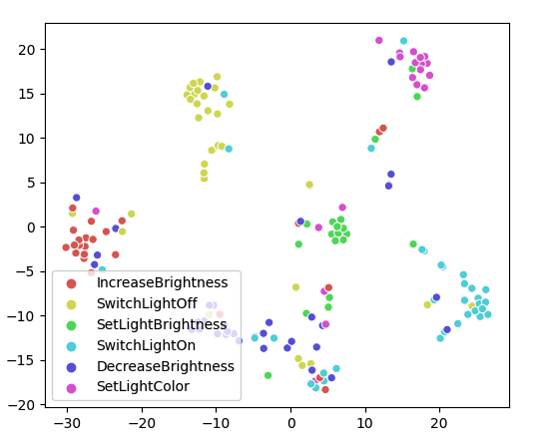}
\caption{\textbf{CMLS Model} - The t-SNE scatter plot of output intents when triplet loss is used. More separable and low-variance clusters are observed compared to the baseline model.}
\label{fig:tsnetriplet2}
\end{minipage}
\end{figure*}

\begin{table}[ht!]
\centering
\begin{tabular}{@{}lr@{}} \toprule
\textbf{Model} & \textbf{Accuracy (\%)}\\ \midrule
Baseline (FSC data, only acoustic, no CMLS) & 96.31 \\
Triplet Loss training (FSC data) & \textbf{97.65}\\
Adversarial training (FSC data) & 96.34 \\ \midrule
Baseline (Snips data, only acoustic, no CMLS) & 70.84\\
SLU Triplet Loss training (Snips data) & \textbf{74.10}\\
SLU Adversarial training (Snips data) & 72.89 \\ \bottomrule
\end{tabular}
\caption{Results on CMLS-based SLU adversarial training}
\label{tab:adv}
\end{table}
Furthermore, we evaluate the predictive power and the false positive rate of our best performing CMLS model (triplet loss) against that of the baseline model.
We plot the confusion matrices for the six intents of the Snips dataset, as predicted by the baseline model (Figure \ref{fig:conf1}) and the CMLS model (Figure \ref{fig:conf2}).
We can see that the baseline model has less predictive power and higher false positive rate. This is especially true for intents that are semantically opposite but acoustically close, such as \emph{IncreaseBrightness} and \emph{DecreaseBrightness}, where we observe the false positive rate of the baseline model ($0.25$) is higher than the CMLS model ($0.11$).
These results further support the superiority of our model in distinguishing intents with similar acoustic features but completely opposite semantic.

We also explore the acoustic embedding representations learned by both for the CMLS and the baseline model. Specifically, we investigate whether the acoustic embeddings pertaining to a particular intent are close to each other or not. For this purpose, we cluster the 768-dimensional hidden acoustic embedding vectors using t-SNE for the test split of the Snips dataset. The vectors are scatter-plotted and color-coded using the ground truth intent, depicted in Figure \ref{fig:tsnetriplet1} (\textit{baseline}) and Figure \ref{fig:tsnetriplet2} (\textit{CMLS with triplet loss}). A visual inspection of these plots reveals that for the CMLS model, the acoustic embedding clusters pertaining to individual intents  are well-separated, having little intra-class variance compared to those obtained from the baseline model. These results further support that our CMLS model better maps audio features to related semantic outputs when compared to models that only use a single modality for the E2E neural SLU task.

\subsubsection{Adversarial Training}
Training our system with the adversarial discriminator network to tie the embedding space, we observe improvements over the baseline, but the triplet loss still remains the strongest candidate for aligning the embedding space, see Table \ref{tab:adv}.
The adversarial training is very sensitive and as outlined in Section \ref{sec:experiments:subsec:training} the optimal tuning parameters are dataset dependent (Table \ref{tab:hypparams}).
Based on these results, it will be interesting to see if a combination of triplet loss and adversarial training complement each other and if the adversarial training will become more robust when applied to a larger dataset.

\section{Conclusion}
\label{sec:conclusion}
In this work, we take a multi-modal view of the E2E SLU model and experiment with a cross-modal latent space (CMLS) setup wherein we learn a shared latent space between the two modalities of the SLU model -- `speech' and `text'.
The CMLS setup helps us leverage modality-specific datasets, that are usually more abundant than E2E SLU datasets, enabling us to achieve a higher SLU performance even on a smaller E2E dataset.
This is done in two ways: using the popular multi-modal losses to explicitly tie the acoustic and text embeddings and by using adversarial training to move the acoustic embeddings closer to the text embeddings.
This is done because the text embeddings are extracted from a semantically powerful BERT-based text encoder, that has been trained on massive amount of textual data and has shown to capture the semantics very well across a variety of tasks.
We show that triplet loss has the best performance and hypothesize that this loss also helps the model recognize off-target instances that are acoustically similar (\textit{sound the same}) but are semantically far apart (e.g `increase', `decrease').
Due the way that it is formulated, it not only minimizes the distance between the target and positive candidates but also maximizes the distance between the target and the audio embeddings that are acoustically close but far apart semantically.
We perform a visual analysis of the acoustic embeddings learned in a cross-modal and non cross-modal way and notice that the cross-modal embeddings not only have have lower false-accept rates but also show better clustering with lower intra-class variance for semantically different intent classes.
Even though adversarial training was not able to beat the triplet loss in a CMLS setup for these datasets, it still performed better than the baseline model that does not have a cross-modal space.

In the future, we plan to combine the adversarial training with the explicit embedding losses in one joint setup and see the combined effects of these methods.
Moreover, we plan to run these experiments on datasets of higher semantic complexity than the ones that we used in this work, and experiment with more sophisticated architectures for the acoustic encoders.
When tying the embeddings, we restricted ourselves to forcing only the acoustic encoder to adjust to the text encoder's latent space because of the limited size of our datasets.
Using larger datasets, having both encoders approximate their outputs to each other may also lead to further performance improvements.
This distributes the stress of approximating the tied space between all of the encoders, that may be beneficial in the case where utterances are similar in one space, but different in another (e.g. acoustically similar utterances with different semantics).

\bibliographystyle{IEEEbib}
\bibliography{strings,refs}
\end{document}